\documentclass{appolb}
\usepackage{graphicx} 
\usepackage{amsmath}
\usepackage{amssymb}
\usepackage{dsfont}

\newcommand{\tr}{\mathop{\mathrm{Tr}}}
\newcommand{\x}{\mathbf{x}}

\newcommand{\nom}{{\nonumber}}

\newcommand{\eref}[1]{Eq.~\eqref{#1}}
\newcommand{\secref}[1]{Sec.~\ref{#1}}
\newcommand{\tabref}[1]{Table~\ref{#1}}
\newcommand{\figref}[1]{Fig.~\ref{#1}}

\begin{document}
\title{Chiral phase transition scenarios from the vector meson extended
Polyakov quark meson model
\thanks{Presented at Excited QCD 2015 (8-14 March 2015, Tatranska Lomnica, Slovakia)}%
} \author{Peter Kov{\'a}cs, Gy{\"o}rgy Wolf \footnote{in collaboration with
    Zsolt Sz{\'e}p} \address{Institute for
    Particle and Nuclear Physics, Wigner Research Centre for Physics,
    Hungarian Academy of Sciences, H-1525 Budapest, Hungary}
  \\
}
\maketitle
\begin{abstract}
Chiral phase transition is investigated in an $SU(3)_L \times SU(3)_R$
symmetric vector meson extended linear sigma model with additional
constituent quarks and Polyakov loops (extended Polyakov quark meson
model). The parameterization of the Lagrangian is done at zero
temperature in a hybrid approach, where the mesons are treated at
tree-level, while the constituent quarks at 1-loop level. The
temperature and baryochemical potential dependence of the two assumed
scalar condensates are calculated from the hybrid 1-loop level
equations of states. The order of the phase transition along the $T=0$
and $\mu_B=0$ axes are determined for various parameterization
scenarios. We find that in order to have a first order phase
transition at $T=0$ as a function of $\mu_B$ a light isoscalar particle is
needed. 
\end{abstract}

  
\section{Introduction}

In \cite{elsm_2013} it was shown through a zero temperature analysis
that $q \bar q$ scalar states, such as the $a_0$, $K^\star_0$, and the
two $f_0$'s are preferred to have masses above $1$~GeV. Similar
results was obtained with $q \bar q$ states in \cite{Chen:2007xr},
while in \cite{Chen:2007xr,Chen:2009gs,Kojo:2008hk} it was shown by
using tetraquarks instead of $q \bar q$ states that the (tetraquark)
scalar masses are in the range $0.6-1.0$~GeV. These results suggests
that the physical states $a_0(1450)$, $K^{\star}_0(1430)$,
$f_0(1370)$, and $f_0(1710)$ (or $f_0(1500)$) are predominantly $\bar
q q$ states, while $a_0(980)$, $K^{\star}_0(800)$, $f_0(500)$, and
$f_0(980)$ are predominantly tetraquark states. However, states with
the same quantum numbers do mix, thus the physical scalar particles
are mixtures of $q \bar q$ and tetraquarks states.  In the current
case of the extended linear $\sigma$ model (EL$\sigma$M) we have only
one scalar nonet, thus we can describe one $a_0$, one $K^\star_0$ and
two $f_0$ (which will be denoted by $f_0^{L/H}$)
particles. Consequently, one of the most interesting question is that
which physical states our fields predominantly are if we investigate
the finite temperature/density behavior of our model additionally to
the zero temperature properties.

The lightest isoscalar ($J^{PC} = 0^{++}$) $q \bar q$ state, $f_0$
(also called $\sigma$) is strongly related to the non-strange
condensate. Since the larger is the sigma mass compared to the mass of
its chiral partner (the pion) the larger is the temperature at which
$m_{f_0}$ approaches $m_\pi$ in the chiral symmetry restoration, one
would expect that a large $m_{f_0}$ mass results in a large
pseudocritical temperature ($T_c$) at zero baryochemical
potentials. On the other hand it is a common expectation that the
chiral phase transition is of first order as a function the
baryochemical potential ($\mu_B$) at $T=0$, and since with increasing
$m_{f_0}$ mass the transition weakens, at some point it is possible
that the transition becomes crossover
\cite{Schaefer:2008hk,Chatterjee:2011jd}. This suggest that for a good
thermodynamic description a small $m_{f_0}$ mass is needed, and
indeed as it turns out our approach supports this requirement.

The paper is organized as follows. In the next subsection we introduce
the model by giving the Lagrangian. In \secref{Sec:param} the
determination of the model parameters is shown, while the description
of the approximation used to calculate the grand potential together
with the field equations are presented in \secref{Sec:field_eq}.


\section{The Model}

The Lagrangian we shall use has the following form:
\begin{align}
  \mathcal{L} & = \tr[(D_{\mu}M)^{\dagger}(D_{\mu}M)] -
  m_{0}^{2}\tr(M^{\dagger}M) - \lambda_{1}[\tr(M^{\dagger} M)]^{2} -
  \lambda_{2}\tr(M^{\dagger}M)^{2} \nom \\
  & + c_{1}(\det M+\det M^{\dagger}) + \tr[H(M+M^{\dagger})]  -
  \frac{1}{4}\tr(L_{\mu\nu}^{2}+R_{\mu\nu}^{2}) \nom \\ 
  & + \tr\left[ \left(\frac{m_{1}^{2}}{2}+\Delta\right)
    (L_{\mu}^{2}+R_{\mu}^{2})\right] + i\frac{g_{2}}{2}(\tr\{L_{\mu\nu}[L^{\mu},L^{\nu}]\} +
  \tr\{R_{\mu\nu}[R^{\mu},R^{\nu}]\}) \nom \\
  & + \frac{h_{1}}{2}\tr(M^{\dagger}M)\tr(L_{\mu} ^{2} + R_{\mu}^{2})
  + h_{2}\tr[(L_{\mu}M)^{2}+(M R_{\mu} )^{2}] \\
  & + 2h_{3}\tr(L_{\mu}M R^{\mu}M^{\dagger}) 
    + \bar{\Psi}\left[i \gamma_{\mu}D^{\mu}-\mathcal{M}\right]\Psi \nom 
\end{align}\label{Eq:Lagr}
The covariant derivatives above are given by
\begin{equation}
  D^{\mu}M = \partial^{\mu}M-i g_{1}(L^{\mu}M-M R^{\mu})-i e
  A_{e}^{\mu}[T_{3},M], \quad D^{\mu}\Psi = \partial^{\mu}\Psi - i
  G^{\mu}\Psi,
\end{equation} 
where $G^{\mu} = g_s G^{\mu}_i T_i,$ with $T_{i} = \lambda_{i}/2$
($i=1,\ldots,8$) denoting the $SU(3)$ group generators given in terms
of the Gell-Mann matrices $\lambda_{i}$. Here $M\equiv M_{S} + M_{PS}$
stands for the scalar\,--\,pseudoscalar fields, $L^{\mu}\equiv V^{\mu} +
A^{\mu}, R^{\mu} \equiv V^{\mu} - A^{\mu}$ for the left and right
handed vector fields (which contain the nonets of vector
($V_{a}^{\mu}$) and axial vector ($A_{a}^{\mu}$) meson fields),
$A_{e}^{\mu}$ is the electromagnetic field, while $G^{\mu}_i$ are the
gluon fields. The field strength tensors are ($Q \in
\{L, R\}$)
\begin{equation}
Q^{\mu\nu} = \partial^{\mu}Q^{\nu} -
ieA_{e}^{\mu}[T_{3},Q^{\nu}] - \left\{\partial^{\nu}Q^{\mu} -
  ieA_{e}^{\nu}[T_{3},Q^{\mu}]\right\}, 
\end{equation} 
while the external fields related to the scalar and vector fields are
$ H = \frac{1}{2}\textnormal{diag} (h_{0N},h_{0N}, \sqrt{2}h_{0S})$,
$\Delta = \textnormal{diag} (\delta_{N}, \delta_{N}, \delta_{S})$ (For
more details on the model see \cite{elsm_2013}).

\section{Setting the Lagrange parameters}
\label{Sec:param}

There are $15$ unknown parameters in \eref{Eq:Lagr}, which will be
present in the field equations at finite temperature and/or density,
namely, $m_0$, $\lambda_1$, $\lambda_2$, $c_1$, $m_1$, $h_1$, $h_2$,
$h_3$, $\delta_N$, $\delta_S$, $\phi_N$, $\phi_S$, $g_F$, $g_1$,
$g_2$. From this set $\delta_N$ can be melted into $m_1$ thus leaving
$14$ unknowns. These parameters are determined similarly as in
\cite{elsm_2013}, that is we calculate values of various masses and
decay widths at tree-level and compare them with the corresponding
experimental value taken from the PDG \cite{PDG} through the $\chi^2$
minimalization process of Ref.~\cite{MINUIT}. It is important to note
that we artificially increased the errors of the PDG to a minimum
level of $5\%$, since we do not expect that our model to be more
precise. Another important points are that now we use a different
anomaly term in \eref{Eq:Lagr} (the term proportional to $c_1$) as
was presented in \cite{elsm_2013}, we fit the total width in case of
$a_0(980)$ instead of the amplitudes, we include the $f_0$ masses and
decay width into the global fit, we take into account of the effects
of the fermion vacuum fluctuations, case of which the expression of
the (pseudo)scalar masses are modified. Moreover, since now we also
included the constituent quarks in the isospin symmetric limit, we use
two additional equations to their tree-level masses with the values
$m_{u,d} = 330$~MeV, and $m_{s} = 500$~MeV.

The scalar meson sector below $2$~GeV contains more physical particles
than we can place into one $q\bar q$ nonet (consisting of $a_0$,
$K_0^{\star}$, $f_0^L$, $f_0^H$), since in nature two $a_0$, two
$K_0^{\star}$ and five $f_0$ particles exist in that energy
range. These particles are the $a_0(980)$ and $a_0(1450)$ (denoted by
$a_0^{1/2}$), the $K_0^{\star}(800)$ and $K_0^{\star}(1430)$, (denoted
by $K_0^{\star\, 1/2}$), the $f_0(500)$ (or $\sigma$), $f_0(980)$,
$f_0(1370)$, $f_0(1500)$ and $f_0(1710)$, (denoted by
$f_0^{1\dots5}$). Accordingly there are $40$ particle assignment
possibilities to pair the physical particles to the members of the
nonet in the model. We performed a $\chi^2$ fit for all the
assignments and ordered them according to their $\chi^2$ values. The
results of the $5$ best solution along with the particle assignments
in two cases (with and without the fermionic vacuum fluctuation) are
shown in \tabref{Tab:chi2}.
\begin{table}[th]
\centering
  \begin{tabular}[c]{|c|c|c|c|c|c|}
\hline
particle assignment & $\chi^2$ & $\chi^2_{\text{red}}$ & particle assignment & $\chi^2$ & $\chi^2_{\text{red}}$\\
\hline
$a_0^1 K_0^{\star\, 2} f_0^{1} f_0^{3}$ & 45.0 & 3.0 & $a_0^1 K_0^{\star\, 2} f_0^{2} f_0^{3}$ & 46.2 & 3.1 \\\hline
$a_0^1 K_0^{\star\, 2} f_0^{1} f_0^{2}$ & 51.7 & 3.4 & $a_0^1 K_0^{\star\, 1} f_0^{1} f_0^{3}$ & 52.7 & 3.5 \\\hline
$a_0^2 K_0^{\star\, 2} f_0^{2} f_0^{3}$ & 51.7 & 3.4 & $a_0^1 K_0^{\star\, 1} f_0^{1} f_0^{2}$ & 54.1 & 3.6 \\\hline
$a_0^1 K_0^{\star\, 2} f_0^{2} f_0^{5}$ & 60.4 & 4.0 & $a_0^1 K_0^{\star\, 2} f_0^{1} f_0^{2}$ & 60.7 & 4.0 \\\hline
$a_0^2 K_0^{\star\, 2} f_0^{3} f_0^{5}$ & 61.3 & 4.1 & $a_0^1 K_0^{\star\, 2} f_0^{1} f_0^{3}$ & 61.7 & 4.1 \\\hline
  \end{tabular}
  \caption{$\chi^2$ and $\chi^2_{\text{red}}\equiv \chi^2/N_\text{dof}$ values ($N_\text{dof} =
    15$) for the first five best solutions of the fit together
    with the particle assignment without (left part) and with (right
    part) the fermionic vacuum fluctuations.}
\label{Tab:chi2}
\end{table}
By a similar fitting procedure in \cite{elsm_2013} we argued that the
best assignment without fitting the $f_0^{L/H}$ is the $a_0^2
K_0^{\star\, 2}$, while we reasoned that  $f_0^{L/H}$ should
correspond to $f_0^{3/5}$. Now it seems that the situation changes
since \tabref{Tab:chi2} one can see that the two best assignment in
case with and without using the fermion vacuum fluctuation are $a_0^1
K_0^{\star\, 2} f_0^{2} f_0^{3}$ and $a_0^1 K_0^{\star\, 2} f_0^{1}
f_0^{3}$, respectively. This suggest that using only zero temperature
quantities in this model is not enough to point out uniquely a single
particle assignment. Thus we investigated the properties of the
different particle assignments at finite temperature/densities as well.
   
\section{Finite temperature field equations }
\label{Sec:field_eq}

In our model there are four order parameters, which are the two chiral
condensates $\phi_N$ (non-strange) and $\phi_S$ (strange) and the two
Polyakov loop variables $\Phi$ and $\bar\Phi$ (For the introduction
of the Polyakov loop variables and their potential see \cite{Polyakov}).
The field equations, which determine the dependence on $T$ and
$\mu_B=3\mu_q$ of the order parameters are given by the minimalization
of the grand canonical potential (see eg. \cite{Kapusta:2006pm}),  
\begin{equation} 
\frac{\partial\Omega_\textnormal{H}}{\partial \phi_N} =
\frac{\partial\Omega_\textnormal{H}}{\partial \phi_S} =
\frac{\partial\Omega_\textnormal{H}}{\partial \Phi} =
\frac{\partial\Omega_\textnormal{H}}{\partial \bar\Phi} = 0,
\label{Eq:field}
\end{equation}
which will result in four coupled equations. In our approach we only
consider vacuum and thermal fluctuations for the constituent quarks
and not for the mesons. The explicit expression for the field
equations can be found in \cite{Fairness_proc_2014} (Eq.~(2)-(5)).

\section{Results and Conclusion}

\begin{figure}[htb]
  \centerline{
    \includegraphics[width=10.5cm]{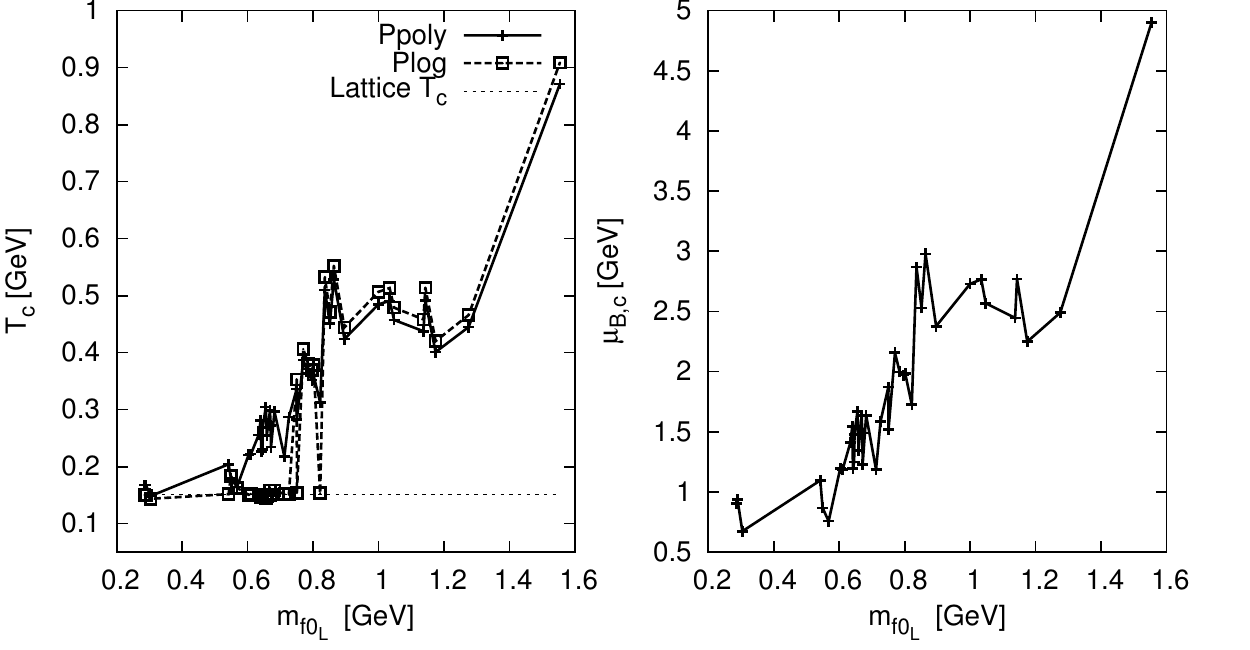}}
  \caption{$T_c$ (left) and $\mu_{B, c}$ (right) values versus
    $m_{f_0^L}$. On the left $T_c$ is shown for two different Polyakov
  potential.}
  \label{Fig:Tc_vs_muf0}
\end{figure}

By solving the field equations \eref{Eq:field} we can investigate the behavior of the
 $\phi_N$, $\phi_S$ order parameters as a function of $T$ at $\mu_B=0$ and as a function
$\mu_B$ at $T=0$. It was calculated on the lattice \cite{aoki_2006}
that at $\mu_B=0$ the value of the pseudocritical temperature $T_C$ should
be $151$~MeV, while it is a common belief that the order of the
transition in $\mu_B$ at $T=0$ should be of first order. On
\figref{Fig:Tc_vs_muf0} the $T_c$ and $\mu_{B, c}$ values for all the 40
assignments are shown\footnote{The lines are only shown to guide the eye.} as a function of $m_{f_0^L}$ (the low lying
isoscalar mass) together with the lattice value for the $T_c$. 
It can be seen that in order to be consistent with the lattice $T_c$
the $m_{f_0^L}$ mass should be below $1$~GeV, which can be correspond
either to $f_0(500)$ or to $f_0(980)$. However if we would like to
have first order phase transition on the $\mu_B$ axis, the $m_{f_0^L}$
mass should be even smaller ($\lessapprox 400$~MeV). Consequently we
investigated the phase boundary for parameterizations with relatively
small $m_{f_0^L}$ masses.

\begin{figure}[htb]
  \centerline{
    \includegraphics[width=10.5cm]{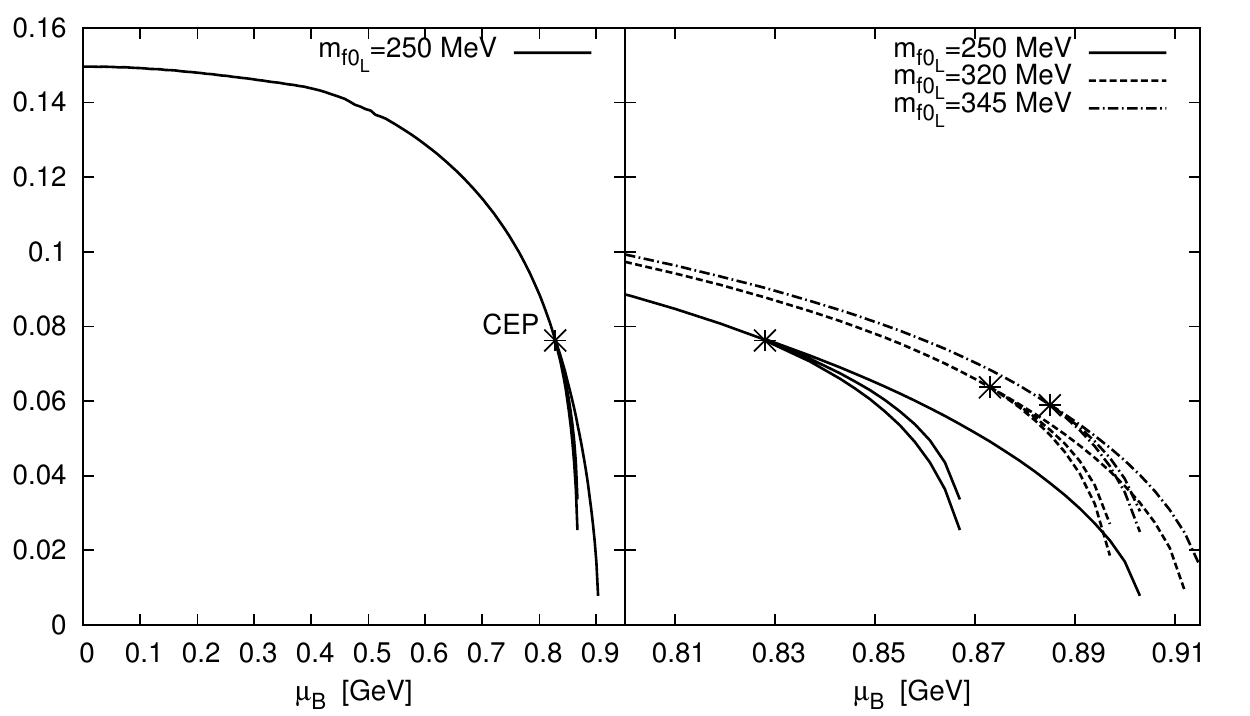}}
  \caption{Phase boundary (left) and CEP variation with $m_{f_0^L}$
    mass (right). The CEP position on the left is at $\mu_{B,\text{CEP}}=828$~MeV, $T_{\text{CEP}}=76$~MeV)}
  \label{Fig:CEP}
\end{figure}

On the left panel of \figref{Fig:CEP} the phase boundary together with
the position of the critical endpoint (CEP) is shown, while on the
right panel the CEP variation with the $m_{f_0^L}$ mass is
presented. 
If we increase the $m_{f_0^L}$ mass above $\approx 350$~MeV
the CEP ceases to exist.

In conclusion we can say that in order to have a good thermodynamic
description within the framework of the current model we must have the
$f_0(500)$ particle in the spectrum. 

\section*{Acknowledgments}

The authors were supported by the Hungarian OTKA fund K109462 and by the
HIC for FAIR Guest Funds of the Goethe University Frankfurt.


\begin{thebibliography}{99}

\bibitem{elsm_2013}  
  D.~Parganlija, P.~Kov\'acs, G.~Wolf, F.~Giacosa and D.~H.~Rischke,
  Phys.\ Rev.\ D {\bf 87}, 014011 (2013).

\bibitem{Chen:2007xr}
  H.~X.~Chen, A.~Hosaka and S.~L.~Zhu,
  Phys.\ Rev.\ D {\bf 76}, 094025 (2007).

\bibitem{Chen:2009gs} 
  H.~X.~Chen, A.~Hosaka, H.~Toki and S.~L.~Zhu,
  Phys.\ Rev.\ D {\bf 81}, 114034 (2010).

\bibitem{Kojo:2008hk} 
  T.~Kojo and D.~Jido,
  Phys.\ Rev.\ D {\bf 78}, 114005 (2008).
  
\bibitem{Schaefer:2008hk} 
  B.~J.~Schaefer and M.~Wagner,
  Phys.\ Rev.\ D {\bf 79}, 014018 (2009).

\bibitem{Chatterjee:2011jd} 
  S.~Chatterjee and K.~A.~Mohan,
  Phys.\ Rev.\ D {\bf 85}, 074018 (2012).

\bibitem {PDG} 
  K.A.~Olive {\it et al.}  (Particle Data Group),
  Chin.\ Phys.\ C {\bf 38}, 090001 (2014)

  
\bibitem{MINUIT} 
  F.~James and M.~Roos,
  Comput.\ Phys.\ Commun.\  {\bf 10} (1975) 343.

\bibitem{Polyakov} 
  G.~Mark{\'o} and Zz.~Sz{\'e}p,
  Phys.\ Rev.\ D {\bf 82}, 065021 (2010);
  S.~Chatterjee and K.~A.~Mohan,
  Phys.\ Rev.\ D {\bf 85}, 074018 (2012);
  H.~Hansen, W.~M.~Alberico, A.~Beraudo, A.~Molinari, M.~Nardi and C.~Ratti,
  Phys.\ Rev.\ D {\bf 75}, 065004 (2007);
  R.~D.~Pisarski,
  Phys.\ Rev.\ D {\bf 62}, 111501 (2000);
  S.~Roessner, C.~Ratti and W.~Weise,
  Phys.\ Rev.\ D {\bf 75}, 034007 (2007).

\bibitem{Fairness_proc_2014}
 P.~Kovács, Z.~Szép and G.~Wolf,
  J.\ Phys.\ Conf.\ Ser.\  {\bf 599}, no. 1, 012010 (2015)
  [arXiv:1501.06426 [hep-ph]].





\bibitem{Kapusta:2006pm}
  J.~I.~Kapusta and C.~Gale,
  {\it Finite-temperature field theory: Principles and applications} (Cambridge University Press, Cambridge, 2006).

\bibitem{aoki_2006} Aoki Y, Fodor Z, Katz S D and Szabo K K 2006
  {\it Phys. Lett.} B {\bf 643} 46































\end{thebibliography}
\end{document}